\documentclass[%
reprint,superscriptaddress,amsmath,amssymb, floatfix,aps,prl,twocolumn]{revtex4-2}
\usepackage{amsmath,amssymb,color,bm,graphicx}
\usepackage[colorlinks=true,linkcolor=blue,citecolor=blue,urlcolor=red]{hyperref}
\usepackage{caption}
\usepackage{subcaption}
\usepackage{orcidlink}
\usepackage[normalem]{ulem}  

\newcommand{\MeV}{\text{MeV}}

\newcommand{\NWA}{\text{NWA}}

\newcommand{\dUrca}{\text{dU}}
\newcommand{\muI}{\Delta\mu}

\begin{document}

\title{Transport properties in binary neutron star mergers: Effect of magnetic field}
\author{Pranjal Tambe~\orcidlink{0000-0003-2293-6953}}
~\email{pranjal.tambe@iucaa.in}
 \affiliation{%
 Inter University Centre for Astronomy and Astrophysics, Ganeshkind, Pune 411007, India
}%
\author{Debarati Chatterjee~\orcidlink{0000-0002-0995-2329}}%
~\email{debarati@iucaa.in}
\affiliation{%
 Inter University Centre for Astronomy and Astrophysics, Ganeshkind, Pune 411007, India
}%

\begin{abstract}
In extreme environments such as binary neutron star mergers,  temperatures as high as $50$ MeV and magnetic fields up to $10^{17}$ G, reach a regime where neutrino transport governs the macroscopic thermodynamic and chemical evolution. Existing merger simulations rely on zero magnetic field neutrino emissivity and opacity, potentially missing critical transport physics in highly magnetized neutron star cores. We present an exact framework for computing charged current Urca emissivity and neutrino opacity at finite temperature and magnetic field. We employ the Nucleon Width Approximation framework to account for the collisional broadening effects dominant in the high-density core. Our calculations demonstrate that extreme magnetic fields significantly enhance charged current neutrino opacity, effectively reducing the mean free path for thermal neutrinos. 
\end{abstract}
\maketitle

\section{Introduction}
\label{sec:intro}

While neutron stars (NS) typically have strong magnetic fields, 
magnetars 
could potentially harbour even higher magnetic fields \cite{Mereghetti:2015asa, Kaspi:2017fwg, Konar:2017kty, Esposito:2018gvp}. Simulations of mergers of binary neutron stars (BNS) indicate that magnetic fields are further amplified during the merger and the resulting remnant can be strongly magnetized \cite{Harding:2006qn, Lorimer:2008se, Kiuchi:2015sga, Ciolfi:2017uak, Ciolfi:2020cpf}. 
In the BNS merger environment, calculating the transport properties is crucial for estimating the dynamics, lifetime, and observable signals of the resulting post-merger remnant \cite{Foucart:2022bth}. Rapid cooling of a newly born neutron star is majorly driven by the emission of neutrinos via Urca processes~\cite{Prakash:1996xs,Yakovlev:2003qy, Janka:2012wk, Pascal:2022qeg}. Despite the fact that it is important to investigate magnetic field effects on transport properties, it has remained an open problem due to the challenges in calculating the effects theoretically.

The dominant channels for neutrino emission are flavor-changing weak interactions. In standard nucleonic matter composed of neutrons, protons, and electrons these are the neutron decay and electron capture processes \cite{Suleiman:2023bdf, Yakovlev:2003qy, Baiko:1998jq}:
\begin{align}
 n&\to p+e^-+\bar\nu_e \quad\text{(nd)} \ , \label{eq:dund-def}\\
 p+e^- &\to n+\nu_e \qquad\quad\text{(ec)} \ . \label{eq:duec-def}
\end{align}

These direct Urca (dU) processes are strongly restricted due to kinematic constraints below a threshold density by momentum conservation requiring proton fraction $x_p>11\%$. In the regions of neutron star where density is insufficient to achieve this threshold proton fraction the dominant reactions are the modified Urca (mU) processes,
\begin{align}
 N+n&\to N+p+e^-+\bar\nu_e \quad\text{(nd)} \ , \label{eq:mund-def}\\
 N+p+e^- &\to N+n+\nu_e \qquad\quad\text{(ec)} \ , \label{eq:muec-def}
\end{align}
where $N=n,p$ is a spectator nucleon. This additional nucleon opens the phase space for the reactions to occur at any density but the requirement of this additional nucleon in the initial and final state suppresses the reaction rates and thus the neutrino emissivity from mU processes by several orders of magnitude compared to the dU processes \cite{Yakovlev:2003qy}. 
The large difference between the emissivities of the dU and mU processes lead to a theoretical distinction between neutron star cooling models. Neutron star models allowing high enough proton fraction to activate the direct Urca processes enable faster cooling causing the surface temperatures to reduce drastically, while models with low proton fraction undergo slower cooling \cite{Yakovlev:2003qy}.

This threshold behavior is however an artifact of the approximation of treating nucleons in the medium as non-interacting particles with infinite lifetime. In a highly dense and strongly interacting medium such as in the core of a neutron star, there are frequent nucleon scattering via strong interaction. Thus the nucleons in this medium have finite lifetime between collisions and thus finite widths in their energy spectra \cite{Suleiman:2023bdf, Sedrakian:2024uma, Alford:2024xfb}. 
The recently developed Nucleon Width Approximation (NWA) framework takes these collisional broadening effects into account. We recently computed the rates of dU reaction processes in presence of magnetic fields, going beyond the Fermi-surface approximation to incorporate thermal effects~\cite{Tambe:2024usx}. We subsequently extended this work using the NWA framework to evaluate the thermal and magnetic field effects on the total Urca rates in~\cite{Tambe:2025evw}.

In this article, we extend the framework to calculate, for the first time, the total neutrino emissivity from Urca processes consistently using the NWA formalism to incorporate the effects of finite temperature and magnetic field. Furthermore, because the transport properties like bulk viscosity depend on the equilibrium composition of nuclear matter, we also compute the neutrino absorption opacity $\left(\nu_e+n\to p + e^-\right)$ and antineutrino absorption opacity $\left( \bar{\nu}_e + p+e^- \to n\right)$ and the corresponding absorption mean free paths. Together the emission and absorption results thus obtained are highly relevant for improving numerical simulations of BNS mergers and core-collapse supernovae, environments where such thermal and magnetic field effects play a significant role.




\begin{figure*}
    \begin{subfigure}{0.47\textwidth}
        \centering
    \includegraphics[width=\textwidth]{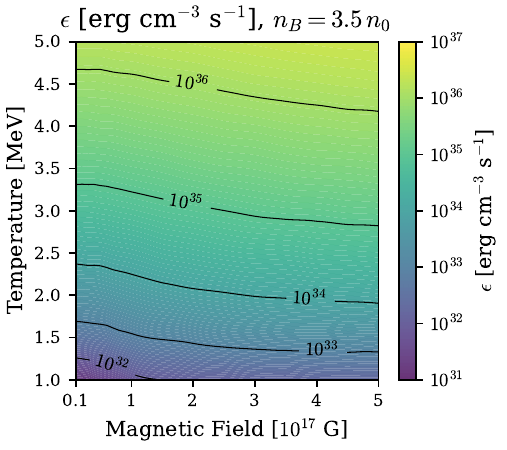} 
    \label{fig:IUF_nb3.5_finiteB}
    \end{subfigure}\hfill
    \begin{subfigure}{0.47\textwidth}
        \centering
    \includegraphics[width=\textwidth]{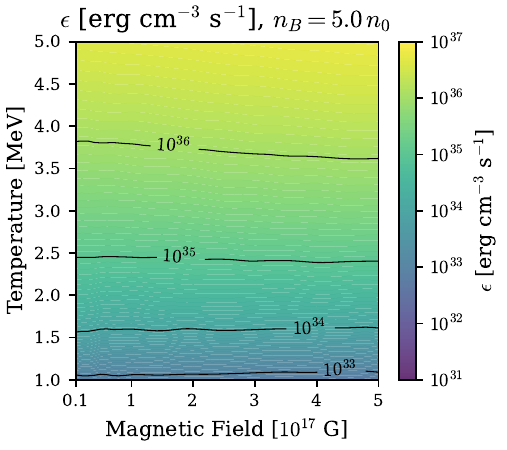} 
    \label{fig:IUF_nb5_finiteB}
    \end{subfigure}
\caption{
Contour plot for emissivity from Urca processes in the NWA formalism for IUF EoS as a function of temperature and magnetic field. }
\label{fig:IUF_finiteB}
\end{figure*}

\begin{figure}
    \centering   \includegraphics[width=0.45\textwidth]{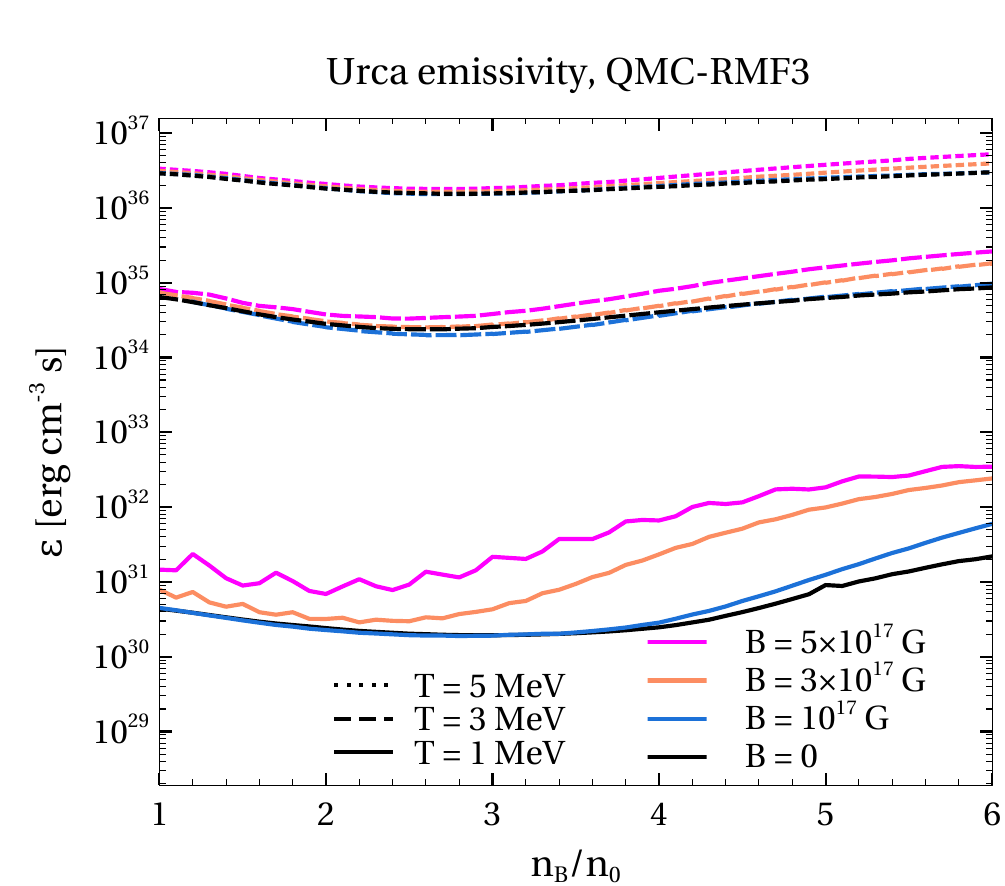}
\caption{
Emissivity from Urca processes in the NWA formalism at three temperatures, $T=1,3,5$ MeV, and magnetic fields $B=10^{17}, 3\times10^{17},$ and $5\times10^{17}$ G using the QMC-RMF3 EoS. }
\label{fig:QMC_finiteB}
\end{figure}

\begin{figure*}
    \begin{subfigure}{0.47\textwidth}
        \centering
    \includegraphics[width=\textwidth]{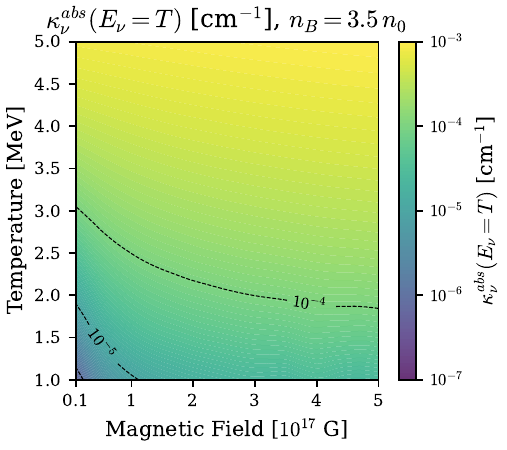} 
    \label{fig:nu_opacity_nb3.5_finiteB}
    \end{subfigure}\hfill
    \begin{subfigure}{0.47\textwidth}
        \centering
    \includegraphics[width=\textwidth]{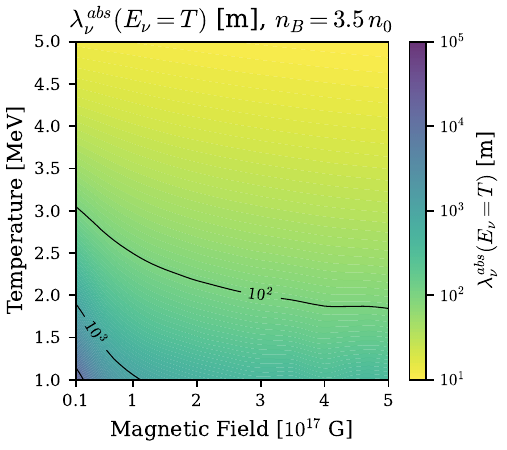} 
    \label{fig:nu_mfp_nb3.5_finiteB}
    \end{subfigure}
\caption{
Contours for neutrino absorption opacity (left) and neutrino absorption mean free path (right) at density $n_B=3.5\,n_0$, below the direct Urca threshold density for IUF EoS, for neutrinos with energy, $E_\nu=T$.}
\label{fig:IUF_neutrino_opacity}
\end{figure*}

\begin{figure*}
    \begin{subfigure}{0.47\textwidth}
        \centering
    \includegraphics[width=\textwidth]{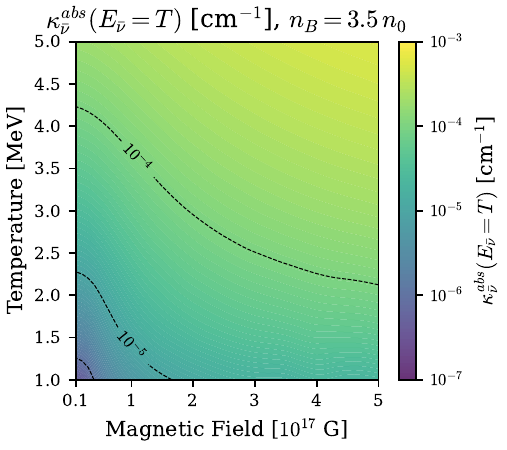} 
    \label{fig:nubar_opacity_nb3.5_finiteB}
    \end{subfigure}\hfill
    \begin{subfigure}{0.47\textwidth}
        \centering
    \includegraphics[width=\textwidth]{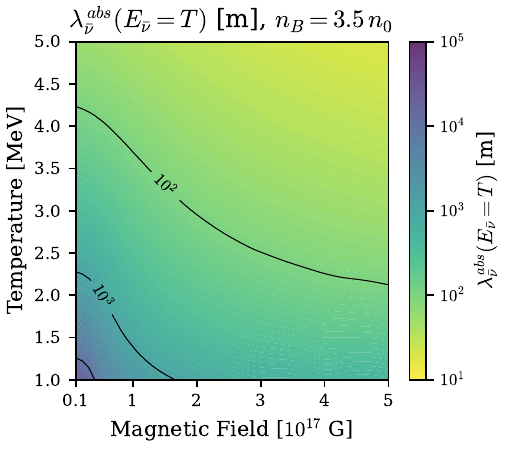} 
    \label{fig:nubar_mfp_nb3.5_finiteB}
    \end{subfigure}
\caption{
Contours for antineutrino absorption opacity (left) and antineutrino absorption mean free path (right) at density $n_B=3.5\,n_0$, below the direct Urca threshold density for IUF EoS, for antineutrinos with energy, $E_\nu=T$.}
\label{fig:IUF_antineutrino_opacity}
\end{figure*}
\section{Formalism}
\label{sec:formalism}

In this work we compute thermal and magnetic field effects on the neutrino transport properties such as emissivity and opacity, above and below the direct Urca threshold density. We 
we only consider nuclear matter composed of neutrons, protons and electrons (npe matter). The extension to include muons is straightforward but would not make a significant difference. 

In this section, we formulate the calculation of the total neutrino emissivity from Urca processes in presence of finite magnetic field and temperature. The antineutrino emissivity from neutron decay direct Urca process in presence of magnetic field is given by \cite{Baiko:1998jq, Tambe:2024usx}:
\begin{align}
    \epsilon^{\dUrca}_\text{nd}=&\frac{G^2(1+3g_A^2)eB}{16\pi^5} \int \,dk_n\, dk_{pz}\, dk_{ez}\, E_{\nu}^3\, k_n\notag\Theta(E_{\nu})\\& \Theta(k_n-\lvert k_{pz}+k_{ez}\rvert)\,\sum_{l,l'} (F^2_{l',l}(u)+F'^2_{l',l-1}(u)) \notag\\& f_n (1-f_p)(1-f_e) \,\label{eq:emissivity_mag},    
\end{align}
where the neutrino energy is obtained from energy conservation relation, $E_{\nu}=E_n-E_p-E_e$. For more details on the definition of the terms and the derivation, we urge the reader to refer to our previous papers~\cite{Tambe:2024usx,Tambe:2025evw}. The Fermi-Dirac distribution functions for constituent particles are given by $f_i=\left[1 + \exp((E_i-\mu_i)/T)\right]^{-1}$. Here $F_{l',l}$ are normalized Laguerre functions given by:
\begin{equation}
    F_{l',l}(u) = \sqrt{\frac{l'!}{l!}} u^{(l-l')/2} e^{{-u}/{2}} L_{l'}^{l-l'}(u)\, = (-1)^{l'-l} F_{l,l'}(u) ,
    \label{eq:lag_f}
\end{equation}
where $L_{l'}^{l-l'}(u)$ are the Laguerre polynomials and $l, l'$ are indices for the electron and proton Landau level numbers~\cite{Tambe:2024usx}. The neutrino emissivity from electron capture process is given by a similar expression by replacing the Fermi-Dirac factors $f_n (1-f_p)(1-f_e)$ by $(1-f_n) f_p f_e$ and the neutrino energy by $E_{\nu}=E_p+E_e-E_n$. 

The total neutrino emissivity in the NWA framework is thus given by \cite{Alford:2024xfb}:
\begin{equation}
    \epsilon^{\NWA} = \int_{-\infty}^{\infty} dm_n\, dm_p\, \epsilon^{\dUrca}(m_n,m_p)\,R_n(m_n)\,R_p(m_p)\, , 
\label{eq:emissivity_NWA}
\end{equation}
where $\epsilon^{\dUrca}$ is the direct Urca emissivity from eq.~(\ref{eq:emissivity_mag}) and the mass-spectral function for nucleon $N\in{n,p}$ is
\begin{equation}
    R_N(m) = \frac{1}{\pi} \frac{W_N/2}{(m - M_N^*)^2 + W_N^2/4}\, .
\label{eq:mass_spectrum}
\end{equation}
Here $M_N^*$ is the in-medium mass of the nucleon and $W_N$ is the nucleon width given by,
\begin{equation}
    W_N = \frac{T^2}{T_W},\quad T_W=5\,\MeV \ .
\end{equation}

Using the same formalism as above for neutrino emissivity we can also compute neutrino absorption opacity \cite{Kumamoto:2024jiq}. The neutrino absorption opacity within the NWA framework is given by,
\begin{align}
    \kappa^{\NWA}_{\nu_e}&=\frac{G^2(1+3g_A^2)eB}{8\pi^3}\int dm_n\, dm_p\, dk_{pz}\, dk_{ez}\, E^*_n\,\notag \\&\Theta(k_n-\lvert k_{pz}+k_{ez}\rvert)\sum_{l,l'}(F^2_{l',l}(u)+F'^2_{l',l-1}(u))\notag \\& f_n(1-f_p)(1-f_e),
    \label{eq:opa_neutrino}
\end{align}
where $E^*_n= \sqrt{k_n^2+m^*_n}$. The antineutrino absorption opacity $\left( \bar{\nu}_e + p+e^- \to n\right)$ within the NWA framework is given by similar expression with Fermi-Dirac factors  $[f_n (1-f_p)(1-f_e)]$ replaced by $[(1-f_n) f_p f_e]$. The neutrino and antineutrino mean free path is just the inverse of their absorption opacities, $\lambda_{\nu}=\kappa_{\nu}^{-1}$.
As in our previous works, we use realistic equations of state (EoS) for nuclear matter which consistently include thermal effects. We do not consider modifications to the EoS due to magnetic field (e.g. Landau quantisation or anomalous magnetic moment) as these effects are not significant for the EoS even for the highest field of $B=5\times10^{17}$ G considered in this work.
We select two realistic finite-temperature relativistic mean field EoS models to describe nuclear matter that satisfy the current nuclear and astrophysical constraints: IUF \cite{Fattoyev:2010mx} and QMC-RMF3 \cite{Alford:2022bpp, Alford:2023rgp}. The IUF EoS has a direct Urca threshold at baryon density $n_{B} = 4.1 n_0$, where $n_0$ is the nuclear saturation density taken to be $0.16$ fm$^{-3}$. The QMC-RMF3 EoS has no direct Urca threshold within the density of interest in this study. \\

\section{Results}
\label{sec:results}

We now compute the total emissivity from Urca processes in the NWA formalism in presence of magnetic field and finite temperature as shown by Fig.~\ref{fig:IUF_finiteB},~\ref{fig:QMC_finiteB} for IUF and QMF-RMF3 EoSs respectively. In npe matter isospin equilibrium is achieved when the neutron decay and electron capture rates are equal, $\Gamma_{nd}=\Gamma_{ec}$. In presence of finite temperature and magnetic field an additional isospin chemical potential $\muI$ is required to achieve this equilibrium as shown in our previous work \cite{Tambe:2025evw}. Thus the true equilibrium condition in presence of finite temperature and magnetic field is $ \mu_n=\mu_p+\mu_e+\muI $. The total Urca emissivity in this work is computed at this true isospin equilibrium condition at all densities as obtained in our previous paper~\cite{Tambe:2025evw}. 

We find that magnetic field enhances the Urca emissivity significantly at low temperatures. We compute emissivity at the true chemical equilibrium condition given by $\mu_n=\mu_p+\mu_e+\Delta\mu$, where $\Delta\mu$ is estimated by equating the neutron decay and electron capture rates \cite{Tambe:2025evw}. 
At low temperatures $T \sim 1-3$ MeV , the Urca emissivity is enhanced by an order of magnitude with increasing magnetic field below the direct Urca threshold density. At higher temperatures $T\sim 5$ MeV, the Urca emissivity increases by a factor of $2$ for the highest magnetic field $B=5\times10^{17}$ G considered in our study. Magnetic field has no significant effect on the Urca emissivity above the direct Urca threshold density.

In Figs.~\ref{fig:IUF_neutrino_opacity}, \ref{fig:IUF_antineutrino_opacity}, we show contour plots for electron neutrino and antineutrino absorption opacity and mean free path due to Urca processes as a function of temperature and magnetic field for soft neutrinos with energy equal to the temperature, $E_{\nu_e,\bar{\nu}_e}=T$. We find that the charged current opacity increases for neutrino and antineutrino in presence of magnetic field. For low temperatures $T< 3$ MeV, the opacities increase by upto $2$ orders of magnitude with an increasing magnetic field. The antineutrino absorption opacity increases by an order of magnitude even at higher temperatures. Correspondingly, we see that the mean free paths due to charged current reactions for both neutrino and antineutrino decreases with increasing magnetic field. For low temperature, $T\sim1$ MeV and low magnetic field $B\sim10^{16}$ G the neutrino and antineutrino absorption mean free paths are larger than the radius of the star, $\lambda_{\nu}>10$ km. The absorption mean free path decreases with increasing magnetic field and temperatures. The combined effects of thermal blurring of Fermi-surfaces, collisional broadening of in-medium nucleons and the magnetic field increase the phase space available for absorption of neutrinos and antineutrinos.  Thus our calculations show that in BNS mergers where magnetic fields as high as $B=5\times10^{17}$ G are reached and the temperatures in the cores are within a few MeV, the neutrino and antineutrino absorption mean free paths can become smaller than the radius of the star. Thus the distribution of neutrinos in BNS merger scenarios will be non-trivial in presence of high magnetic fields even at lower temperatures. This will modify the equilibrium composition of matter in the merger scenarios and also affect the viscous damping of density oscillations.



\section{Discussions}
\label{sec:discussions}

In this work, we extended our recently developed formalism to obtain weak interaction rates including thermal and magnetic field effects within the NWA framework to calculate transport properties consistently in the post-merger scenario.
We found that magnetic field enhances the Urca emissivity significantly at low temperatures below the direct Urca threshold density. This will lead to a faster cooling of the highly magnetized post-merger remnant core.

We also computed the absorption opacities for neutrinos and antineutrinos in presence of magnetic field incorporating thermal and collisional broadening effects. We saw that the opacities are enhanced with increasing magnetic field at low temperatures. Thus the absorption mean free path decreases by orders of magnitude at low temperatures with increasing magnetic fields, becoming smaller than the radius of the star even at low temperatures. This will modify the equilibrium composition of matter in the core of the post-merger. This will also modify the bulk viscous damping of oscillations in BNS mergers. Therefore it becomes crucial to incorporate the magnetic field effects on the neutrino opacities in simulations of BNS mergers.


Our calculations show that magnetic field effects can have important consequences on the calculations of transport properties of the BNS post-merger remnant, where such magnetic fields and temperatures may be encountered. The results of this work can be followed up with more detailed numerical simulations, incorporating more sophisticated magnetic field and thermal evolution as well as neutrino transport. Thermal evolution of the remnant determines its stability as well as the threshold mass before collapse to a black hole (see e.g.~\cite{Weih:2017mcw, Koppel:2019pys, Nunna:2020jzt, Lucca:2019ohp}), with consequences for multi-messenger astrophysical observations following the detection of gravitational waves from a merger event. 
\\

\section{Acknowledgements}
P.T. acknowledges Mark Alford, Alexander Haber, Liam Brodie for insightful discussions. P.T. and D.C. acknowledge the usage of IUCAA HPC computing facility for computations performed in this paper.
\bibliography{bibliography}

@article{Kiuchi:2015sga,
    author = "Kiuchi, Kenta and Cerd\'a-Dur\'an, Pablo and Kyutoku, Koutarou and Sekiguchi, Yuichiro and Shibata, Masaru",
    title = "{Efficient magnetic-field amplification due to the Kelvin-Helmholtz instability in binary neutron star mergers}",
    eprint = "1509.09205",
    archivePrefix = "arXiv",
    primaryClass = "astro-ph.HE",
    doi = "10.1103/PhysRevD.92.124034",
    journal = "Phys. Rev. D",
    volume = "92",
    number = "12",
    pages = "124034",
    year = "2015"
}

@article{Ciolfi:2017uak,
    author = "Ciolfi, Riccardo and Kastaun, Wolfgang and Giacomazzo, Bruno and Endrizzi, Andrea and Siegel, Daniel M. and Perna, Rosalba",
    title = "{General relativistic magnetohydrodynamic simulations of binary neutron star mergers forming a long-lived neutron star}",
    eprint = "1701.08738",
    archivePrefix = "arXiv",
    primaryClass = "astro-ph.HE",
    doi = "10.1103/PhysRevD.95.063016",
    journal = "Phys. Rev. D",
    volume = "95",
    number = "6",
    pages = "063016",
    year = "2017"
}

@article{Ciolfi:2020cpf,
    author = "Ciolfi, Riccardo",
    title = "{The key role of magnetic fields in binary neutron star mergers}",
    eprint = "2003.07572",
    archivePrefix = "arXiv",
    primaryClass = "astro-ph.HE",
    doi = "10.1007/s10714-020-02714-x",
    journal = "Gen. Rel. Grav.",
    volume = "52",
    number = "6",
    pages = "59",
    year = "2020"
}

@article{Alford:2023rgp,
    author = "Alford, Mark G. and Brodie, Liam and Haber, Alexander and Tews, Ingo",
    title = "{Tabulated equations of state from models informed by chiral effective field theory}",
    eprint = "2304.07836",
    archivePrefix = "arXiv",
    primaryClass = "nucl-th",
    reportNumber = "LA-UR-23-23837",
    doi = "10.1088/1402-4896/ad03c8",
    journal = "Phys. Scripta",
    volume = "98",
    number = "12",
    pages = "125302",
    year = "2023"
}

@article{Harding:2006qn,
    author = "Harding, Alice K. and Lai, Dong",
    title = "{Physics of Strongly Magnetized Neutron Stars}",
    eprint = "astro-ph/0606674",
    archivePrefix = "arXiv",
    doi = "10.1088/0034-4885/69/9/R03",
    journal = "Rept. Prog. Phys.",
    volume = "69",
    pages = "2631",
    year = "2006"
}

@article{Lorimer:2008se,
    author = "Lorimer, D. R.",
    title = "{Binary and Millisecond Pulsars}",
    eprint = "0811.0762",
    archivePrefix = "arXiv",
    primaryClass = "astro-ph",
    doi = "10.12942/lrr-2008-8",
    journal = "Living Rev. Rel.",
    volume = "11",
    pages = "8",
    year = "2008"
}

@article{Baiko:1998jq,
    author = "Baiko, D. A. and Yakovlev, D. G.",
    title = "{Direct {Urca} process in strong magnetic fields and neutron star cooling}",
    eprint = "astro-ph/9812071",
    archivePrefix = "arXiv",
    journal = "Astron. Astrophys.",
    volume = "342",
    pages = "192--200",
    year = "1999"
}

@article{Kaspi:2017fwg,
    author = "Kaspi, Victoria M. and Beloborodov, Andrei",
    title = "{Magnetars}",
    eprint = "1703.00068",
    archivePrefix = "arXiv",
    primaryClass = "astro-ph.HE",
    doi = "10.1146/annurev-astro-081915-023329",
    journal = "Ann. Rev. Astron. Astrophys.",
    volume = "55",
    pages = "261--301",
    year = "2017"
}

@article{Konar:2017kty,
    author = "Konar, Sushan",
    title = "{Magnetic Fields of Neutron Stars}",
    eprint = "1709.07106",
    archivePrefix = "arXiv",
    primaryClass = "astro-ph.HE",
    doi = "10.1007/s12036-017-9467-4",
    journal = "J. Astrophys. Astron.",
    volume = "38",
    pages = "47",
    year = "2017"
}

@article{Alford:2022bpp,
    author = "Alford, Mark G. and Brodie, Liam and Haber, Alexander and Tews, Ingo",
    title = "{Relativistic mean-field theories for neutron-star physics based on chiral effective field theory}",
    eprint = "2205.10283",
    archivePrefix = "arXiv",
    primaryClass = "nucl-th",
    reportNumber = "LA-UR-22-23972",
    doi = "10.1103/PhysRevC.106.055804",
    journal = "Phys. Rev. C",
    volume = "106",
    number = "5",
    pages = "055804",
    year = "2022"
}

@misc{Alford:2024xfb,
    author = "Alford, Mark G. and Haber, Alexander and Zhang, Ziyuan",
    title = "{Beyond modified {Urca}: the nucleon width approximation for flavor-changing processes in dense matter}",
    eprint = "2406.13717",
    archivePrefix = "arXiv",
    primaryClass = "nucl-th",
    month = "6",
    year = "2024"
}

@article{Foucart:2022bth,
    author = "Foucart, Francois",
    title = "{Neutrino transport in general relativistic neutron star merger simulations}",
    eprint = "2209.02538",
    archivePrefix = "arXiv",
    primaryClass = "astro-ph.HE",
    doi = "10.1007/s41115-023-00016-y",
    journal = "Liv. Rev. Comput. Astrophys.",
    volume = "9",
    number = "1",
    pages = "1",
    year = "2023"
}

@misc{Tambe:2024usx,
    author = "Tambe, Pranjal and Chatterjee, Debarati and Alford, Mark and Haber, Alexander",
    title = "{Effect of Magnetic Fields on Urca Rates in Neutron Star Mergers}",
    eprint = "2409.09423",
    archivePrefix = "arXiv",
    primaryClass = "nucl-th",
    month = "9",
    year = "2024"
}

@article{Esposito:2018gvp,
    author = "Esposito, P. and Rea, N. and Israel, G. L.",
    editor = "Belloni, Tomaso M. and M\'endez, Mariano and Zhang, Chengmin",
    title = "{Magnetars: a short review and some sparse considerations}",
    eprint = "1803.05716",
    archivePrefix = "arXiv",
    primaryClass = "astro-ph.HE",
    doi = "10.1007/978-3-662-62110-3_3",
    journal = "Astrophys. Space Sci. Libr.",
    volume = "461",
    pages = "97--142",
    year = "2020"
}

@article{Mereghetti:2015asa,
    author = "Mereghetti, Sandro and Pons, Jose' and Melatos, Andrew",
    title = "{Magnetars: Properties, Origin and Evolution}",
    eprint = "1503.06313",
    archivePrefix = "arXiv",
    primaryClass = "astro-ph.HE",
    doi = "10.1007/s11214-015-0146-y",
    journal = "Space Sci. Rev.",
    volume = "191",
    number = "1-4",
    pages = "315--338",
    year = "2015"
}

@misc{Tambe:2025evw,
    author = "Tambe, Pranjal and Chatterjee, Debarati and Alford, Mark and Haber, Alexander",
    title = "{Thermal and Magnetic effects on Bulk Viscosity in Binary Neutron Star Mergers}",
    eprint = "2510.09104",
    archivePrefix = "arXiv",
    primaryClass = "nucl-th",
    month = "10",
    year = "2025"
}

@article{Suleiman:2023bdf,
    author = "Suleiman, Lami and Oertel, Micaela and Mancini, Marco",
    title = "{Modified Urca neutrino emissivity at finite temperature}",
    eprint = "2308.09819",
    archivePrefix = "arXiv",
    primaryClass = "nucl-th",
    doi = "10.1103/PhysRevC.108.035803",
    journal = "Phys. Rev. C",
    volume = "108",
    number = "3",
    pages = "035803",
    year = "2023"
}

@article{Yakovlev:2003qy,
    author = "Yakovlev, D. G. and Gnedin, Oleg Y. and Kaminker, A. D. and Levenfish, K. P. and Potekhin, Alexander Y.",
    title = "{Neutron star cooling: Theoretical aspects and observational constraints}",
    eprint = "astro-ph/0306143",
    archivePrefix = "arXiv",
    doi = "10.1016/j.asr.2003.07.020",
    journal = "Adv. Space Res.",
    volume = "33",
    number = "4",
    pages = "523--530",
    year = "2004"
}

@article{Janka:2012wk,
    author = "Janka, Hans-Thomas",
    title = "{Explosion Mechanisms of Core-Collapse Supernovae}",
    eprint = "1206.2503",
    archivePrefix = "arXiv",
    primaryClass = "astro-ph.SR",
    doi = "10.1146/annurev-nucl-102711-094901",
    journal = "Ann. Rev. Nucl. Part. Sci.",
    volume = "62",
    pages = "407--451",
    year = "2012"
}

@article{Prakash:1996xs,
    author = "Prakash, Madappa and Bombaci, Ignazio and Prakash, Manju and Ellis, Paul J. and Lattimer, James M. and Knorren, Roland",
    title = "{Composition and structure of protoneutron stars}",
    eprint = "nucl-th/9603042",
    archivePrefix = "arXiv",
    reportNumber = "SUNY-NTG-96-11, NUC-MINN-93-23-T",
    doi = "10.1016/S0370-1573(96)00023-3",
    journal = "Phys. Rept.",
    volume = "280",
    pages = "1--77",
    year = "1997"
}

@article{Pascal:2022qeg,
    author = "Pascal, A. and Novak, J. and Oertel, M.",
    title = "{Proto-neutron star evolution with improved charged-current neutrino{\textendash}nucleon interactions}",
    eprint = "2201.01955",
    archivePrefix = "arXiv",
    primaryClass = "nucl-th",
    doi = "10.1093/mnras/stac016",
    journal = "Mon. Not. Roy. Astron. Soc.",
    volume = "511",
    number = "1",
    pages = "356--370",
    year = "2022"
}

@article{Fattoyev:2010mx,
    author = "Fattoyev, F. J. and Horowitz, C. J. and Piekarewicz, J. and Shen, G.",
    title = "{Relativistic effective interaction for nuclei, giant resonances, and neutron stars}",
    eprint = "1008.3030",
    archivePrefix = "arXiv",
    primaryClass = "nucl-th",
    doi = "10.1103/PhysRevC.82.055803",
    journal = "Phys. Rev. C",
    volume = "82",
    pages = "055803",
    year = "2010"
}

@article{Sedrakian:2024uma,
    author = "Sedrakian, Armen",
    title = "{Short-Range Correlations and Urca Process in Neutron Stars}",
    eprint = "2406.16183",
    archivePrefix = "arXiv",
    primaryClass = "nucl-th",
    doi = "10.1103/PhysRevLett.133.171401",
    journal = "Phys. Rev. Lett.",
    volume = "133",
    number = "17",
    pages = "171401",
    year = "2024"
}

@article{Kumamoto:2024jiq,
    author = "Kumamoto, Mia and Welch, Catherine",
    title = "{Effects of Landau quantization on neutrino emission and absorption}",
    eprint = "2412.02925",
    archivePrefix = "arXiv",
    primaryClass = "nucl-th",
    reportNumber = "INT-PUB-24-058, INT-PUB-24-041",
    doi = "10.1103/PhysRevD.111.063009",
    journal = "Phys. Rev. D",
    volume = "111",
    number = "6",
    pages = "063009",
    year = "2025"
}

@article{Nunna:2020jzt,
    author = "Nunna, Krishna Prakash and Banik, Sarmistha and Chatterjee, Debarati",
    title = "{Signatures of strangeness in neutron star merger remnants}",
    eprint = "2002.07538",
    archivePrefix = "arXiv",
    primaryClass = "astro-ph.HE",
    doi = "10.3847/1538-4357/ab8f2c",
    journal = "Astrophys. J.",
    volume = "896",
    number = "2",
    pages = "109",
    year = "2020"
}

@article{Weih:2017mcw,
    author = "Weih, Lukas R. and Most, Elias R. and Rezzolla, Luciano",
    title = "{On the stability and maximum mass of differentially rotating relativistic stars}",
    eprint = "1709.06058",
    archivePrefix = "arXiv",
    primaryClass = "gr-qc",
    doi = "10.1093/mnrasl/slx178",
    journal = "Mon. Not. Roy. Astron. Soc.",
    volume = "473",
    number = "1",
    pages = "L126--L130",
    year = "2018"
}

@article{Koppel:2019pys,
    author = {K{\"o}ppel, Sven and Bovard, Luke and Rezzolla, Luciano},
    title = "{A General-relativistic Determination of the Threshold Mass to Prompt Collapse in Binary Neutron Star Mergers}",
    eprint = "1901.09977",
    archivePrefix = "arXiv",
    primaryClass = "gr-qc",
    doi = "10.3847/2041-8213/ab0210",
    journal = "Astrophys. J. Lett.",
    volume = "872",
    number = "1",
    pages = "L16",
    year = "2019"
}

@article{Lucca:2019ohp,
    author = "Lucca, Matteo and Sagunski, Laura",
    title = "{The lifetime of binary neutron star merger remnants}",
    eprint = "1909.08631",
    archivePrefix = "arXiv",
    primaryClass = "astro-ph.HE",
    reportNumber = "TTK-19-35",
    doi = "10.1016/j.jheap.2020.04.003",
    journal = "JHEAp",
    volume = "27",
    pages = "33--37",
    year = "2020"
}
\end{document}